# Brillouin Gain Microscopy

Roni Shaashoua,[1] Lir Kasuker,[2] Mor Kishner,[2] Tal Levy,[2] Barak Rotblat,[2,3] Anat Ben-Zvi[2] & Alberto Bilenca[1,4,*]

[1]Biomedical Engineering Department, Ben-Gurion University of the Negev, 1 Ben Gurion Blvd, Be'er-Sheva 8410501, Israel

[2]Department of Life Sciences, Ben-Gurion University of the Negev, 1 Ben Gurion Blvd, Be'er-Sheva 8410501, Israel

[3]The National Institute for Biotechnology in the Negev, Ben-Gurion University of the Negev, 1 Ben Gurion Blvd, Be'er-Sheva 8410501, Israel

[4]Ilse Katz Institute for Nanoscale Science and Technology, Ben-Gurion University of the Negev, 1 Ben Gurion Blvd, Be'er-Sheva 8410501, Israel

e-mail: *bilenca@bgu.ac.il

**Abstract**

Optical imaging with mechanical contrast is critical for material and biological discovery since it allows contactless light-radiation force-excitation within the sample, as opposed to traditional mechanical imaging. Whilst optical microscopy based on stimulated Brillouin scattering (SBS) enables mechanical imaging of materials and living biological systems with high spectrospatial resolution, its temporal resolution remains limited. Here, we develop Brillouin gain microscopy (BGM) with a 200-fold higher temporal resolution by detecting the Brillouin gain at a mechanically contrasting frequency in the sample with high sensitivity. Using BGM, we demonstrate mechanical imaging of materials, living organisms and cells at high spectro-spatiotemporal resolution.

**Introduction**

Optical mechanical imaging techniques are complementary to atomic force microscopy (AFM), which has traditionally been applied for mechanical research and discovery in many fields including biology and material sciences. Whilst AFM allows for two-dimensional mechanical measurements on the sample surface at nanometer scale[1,2], optical mechanical microscopy facilitates cross-sectional imaging inside the sample along all three dimensions at (sub)micrometer scale. Among the established optical techniques used for imaging materials and living biological systems with mechanical contrast are optical coherence elastography (OCE)[3-5] and second harmonic generation (SHG) microscopy[6-9]. OCE is based on the application of a mechanical load to the sample and measurement of the subsequent deformation by speckle tracking or phase-sensitive detection. Whereas OCE achieves ~5-15 μm resolution and impressively fast acquisition times (several seconds to tens of minutes, depending on the imaged field-of-view and pixel density), it involves intensive image acquisition and processing, and the use of specialized instrumentation for external mechanical stimulation and force sensing. SHG microscopy circumvents the need for an external mechanical load (i.e., is all-optical) while offering high spatiotemporal resolution, however, its applicability is limited to a small number of structural proteins.

Stimulated Brillouin scattering (SBS) microscopy is emerging as a promising method for all-optical mechanical imaging with subcellular spatial resolution, high mechanical specificity (~100-MHz spectral resolution), and high sensitivity (fractional intensity of ~$10^{-5}$ in 100 μs in biological settings)[10]. In SBS microscopy, the interaction of the excitation light at the pump and Stokes frequencies, $\omega_P$ and $\omega_S$, with a propagating coherent acoustic wave (acoustic phonon) at a frequency $\Omega_B = \omega_P - \omega_S$ of acoustic vibration in the sample results in an increase of the Stokes beam intensity (Fig. 1, A and B). By scanning the difference frequency $\Omega(t) = \omega_P - \omega_S(t)$ over $\Omega_B$ in time, the so-called stimulated Brillouin gain (SBG) spectrum is obtained (Fig. 1B in gray line). To measure the SBG at $\Omega$, we use a high frequency phase-sensitive detection scheme by modulating the pump beam intensity at 1.1 MHz and detecting the SBG in the Stokes beam with a lock-in amplifier. Suppression of the unwanted pump back-reflections from the sample into the detector is accomplished by cleaning the optical noise in the forward-going, unmodulated pump beam with a Bragg grating and blocking of the back-reflected, and the modulated pump beam by an atomic filter (Fig. 1C). For a known refractive index of the sample, the Brillouin frequency shift, the Brillouin linewidth, and the peak Brillouin gain of the SBG spectrum are directly related to the acoustic speed, acoustic attenuation, and mass density in the sample. Using this acoustic information, the complex longitudinal modulus of the material constituents in the sample can be determined at gigahertz frequencies. Spontaneous Brillouin scattering microscopy based on light diffraction from thermal acoustic phonons can also be used for all-optical mechanical imaging, but it trades-off acquisition time for spectral and/or spatial resolution[11-21]. Despite the substantial improvements enabled by SBS microscopy for optical mechanical imaging, its temporal resolution is still limited to a few tens of milliseconds under biocompatible excitation power.

In general, Brillouin microscopes are supposed to measure the entire Brillouin spectrum in each image pixel to extract $\Omega_B$ and therefore to obtain mechanical information about the sample. In SBS microscopy, this requires the sweeping of the Stokes frequency $\omega_S$—a time-consuming process that significantly limits the temporal resolution. For achieving high speed imaging, we develop here Brillouin gain microscopy (BGM), which measures the Brillouin gain at a single, mechanically contrasting frequency of the sample $\Omega_C$ similar to stimulated Raman scattering (SRS) microscopy[22-24]. For example, in a sample with a single material constituent, $\Omega_C$ can be chosen to be at the Brillouin frequency shift of the material $\Omega_B$ (Fig. 1B). Due to the detection of unwanted pump back-reflections from the sample, there is a significant drift in the baseline level of the SBG spectra acquired by SBS microscopy at different locations in the sample (Fig. 1D). This baseline drift fundamentally impedes the accurate measurement of the Brillouin gain at $\Omega_C$. We solve this problem in BGM by designing an optimized optical filter based on a greatly simplified, free-running, temperature-controlled solid etalon (Methods, Note S1, and Fig. S1) that improves the suppression of the integrated optical noise power in the forward-going, unmodulated pump beam by ~20 dB with an insertion loss of ~0.75 dB (Fig. 1E). This optimization is necessary to straighten the baseline with minimal pump power loss, enabling the accurate and precise measurement of the Brillouin gain at $\Omega_C$ (Fig. S2). While our previous SBS microscope could not accurately measure the Brillouin gain at $\Omega_C/2\pi = \Omega_B^{water}/2\pi = 5.03$ GHz across a glass-water interface, the new Brillouin gain microscopy (BGM) system successfully acquired this data with temporal resolution of 100 μs—200-fold smaller than the pixel-dwell time required to measure a whole SBG spectrum (Fig. 1F and Fig. S3). Effectively corrected baselines were also obtained in more complex samples, including living *C. elegans* nematodes and NIH/3T3 cells (Fig. S4). The overall power on the sample was ~228 mW. Owing to the significantly reduced light exposure during data acquisition, phototoxicity is substantially lower than with our previous SBS microscope[10].

As the first application, we used BGM for material imaging of phosphate-buffered saline (PBS) on polyacrylamide (PAA) gel (Fig. 2A and Methods). The SBG spectrum of PBS has a peak at $\Omega_B^{PBS}/2\pi = 5.06$ GHz, whereas PAA has a Brillouin peak at $\Omega_B^{PAA}/2\pi = 5.39$ GHz (Fig. 2B), attributable to the stiffer nature of PAA compared with PBS. Figure 2, C and D show the BGM images obtained by the acoustic vibration frequencies of the PBS and PAA, respectively ($\Omega_B^{PBS}$ and $\Omega_B^{PAA}$). Contrast using $\Omega_B^{PBS}$ highlights the top layer of the sample (PBS) with positive contrast and the bottom layer (PAA) with negative contrast. The opposite occurs by tuning in to $\Omega_B^{PAA}$. The merged image of the PBS and PAA channels clearly shows the two layers of the material system (Fig. 2E). While the use of either $\Omega_B^{PBS}$ or $\Omega_B^{PAA}$ provides a significant difference between the materials, setting the mechanically contrasting frequency $\Omega_C$ to $\Omega_B^{PAA}$ offers a higher contrast between PBS and PAA than $\Omega_B^{PBS}$ (Fig. 2F). This result is reasoned by the asymmetric aperture-induced spectral broadening of Brillouin peaks[25].

Next, we present in vivo BGM imaging of the head of a *C. elegans* nematode, which serves as an important multicellular model organism in biological research (Fig. 3 and Methods). The pharynx of the nematode exhibits a Brillouin peak at $\Omega_B^{pharynx}/2\pi = 5.53$ GHz, whereas the surrounding tissue of the nematode's head has a Brillouin peak at $\Omega_B^{surrounding\ tissue}/2\pi = 5.3$ GHz, implying that the pharynx is stiffer than the surrounding tissue probably due to the muscular nature of the pharynx[10] (Fig. 3A). The tissue around the pharynx can be seen with positive contrast and the pharynx with negative contrast when tuned in to the 5.3-GHz acoustic vibration frequency (Fig. 3B). The corpus, isthmus, and terminal bulb of the pharynx are clearly exposed together with the pharyngeal lumen and grinder. Contrast coming from the pharynx channel (5.53 GHz) is also available (Fig. 3C) although strong back-reflected pump signals off interfaces of the pharynx and the surrounding tissue appear to somewhat increase the SBG artificially at these locations. Figure 3D illustrates the dual-color overlay of the surrounding tissue and pharynx channels showing the stiffer pharynx in cyan and the softer surrounding tissue in red. Another useful mechanically contrasting frequency $\Omega_C$ is situated near the middle of the slope of the SBG spectrum of the pharynx ($\Omega_C/2\pi = 5.7$ GHz as depicted in Fig. 3A). At this frequency, changes in the Brillouin frequency shift are efficiently converted to SBG variations owing to mechanically different constituents of the sample. The resultant BGM image shows high contrast between the pharynx and the surrounding tissue regions (Fig. 3E), as explained by the greater visibility between the mean SBGs of these two regions at 5.7 GHz compared to 5.3 GHz and 5.53 GHz (Fig. 3F).

Another ability of BGM is to image biological cells with mechanical contrast at subcellular resolution despite the closely overlapping Brillouin bands of different components in the cell (Fig. 4). Figure 4A displays the SBG spectra of the cell nucleoplasm and nucleolus in a living NIH/3T3 cell (Methods), which is a widely used fibroblast cell line in biological studies. These cell components show Brillouin peaks at $\Omega_B^{nucleoplasm}/2\pi$ = 5.25 GHz and $\Omega_B^{nucleolus}/2\pi$ = 5.4 GHz (i.e., only 150 MHz frequency difference), suggesting a stiffer nucleolus. By tuning in to $\Omega_B^{nucleoplasm}$, contrast from regions outside the nucleoli appeared so that the nucleoli could be identified (Fig. 4B). The nucleoli are hardly seen when tuned in to the nucleolus Brillouin peak at $\Omega_B^{nucleolus}$ due to its significant overlap with the nucleoplasm band (Fig. 4, A and C). The dual-color overlay image (Fig. 4D) shows the two nucleoli, the nucleoplasm (light gray region), and the cytoplasm (dark gray region). Figure 4E shows that high contrast is available between these three regions when the mechanically contrasting frequency $\Omega_C$ is tuned near the middle of the slope of the SBG spectrum of the nucleolus ($\Omega_C/2\pi$ = 5.7 GHz as illustrated in Fig. 4A). Whereas the contrast coming from the acoustic vibration frequencies 5.25 GHz, 5.4 GHz, and 5.7 GHz yields statistically different cell regions, visibility is best at 5.7 GHz (Fig. 4F).

BGM overcomes the frequency-space-time tradeoff challenge in all-optical mechanical imaging to provide SBS microscopy with intrinsic mechanical contrast at high spectro-spatiotemporal resolution. BGM can now be used in various applications in the material and biological sciences, including those involving living organisms and cells. As the low-frequency counterpart of SRS microscopy, BGM is likely to become an increasingly powerful tool in material and biological research and discovery.

**Methods**

*Temperature-controlled solid etalon*

The solid etalon (LightMachinery) was mounted in a custom-made aluminum mount, placed on a 5-axis stage (Thorlabs). A thermoelectric cooler was positioned on top of the mount, above a conductive tape, with a heat sink on top of it (Fig. S1A). The mount was covered with insulating pads. A thermistor probe measured the temperature of the etalon directly through a hole on the side of the mount. The pump beam traveled through the etalon twice using a prism (Foctek Photonics) and a PID temperature controller (Thorlabs) kept the temperature of the etalon stable. The device was covered by an environmentally insulating box. The transmission response and frequency stability of the two-pass etalon was measured showing a transmission peak spectral separation, the so-called free spectral range (FSR), of 15.14 GHz, a peak linewidth of 228 MHz in full-width at half-maximum, resulting in a finesse of ~66, and frequency stability better than ~13 MHz over one hour (Fig. S1, B and C). The measured maximum transmission of the two-pass etalon was ~82%.

*Preparation of the two-layer sample of phosphate-buffered saline (PBS) and polyacrylamide (PAA) gel*

We produced two-layer samples consisting of a PBS layer overlying a PAA gel layer (Fig. 2). The PAA-PBS sample holder comprised two 0.15-mm thick glass coverslips, 25 mm and 18 mm in diameter, spaced at 0.36 mm using adhesive spacers (Grace Bio-Labs SecureSeal). A ~120-μm thick PAA gel layer was prepared on the bottom, 25-mm diameter coverslip by mixing 450 µl of 40% acrylamide (Bio-Rad), 200 µl of 2% bis-acrylamide (Bio-Rad), and 350 µl PBS (Biological Industries). 3 µl tetramethylethylenediamine (Merck) and 10 µl of 10% ammonium persulfate (Merck) were added to catalyze the acrylamide and bisacrylamide polymerization. A rain-X-coated coverslip was placed on top of the PAA mixture to flatten the surface of the gel. After the gel was solidified, the rain-X-coated coverslip was removed. The gel-coated coverslip was kept soaked in PBS at 4°C. For BGM imaging, the space between the two glass coverslips of the sample holder was filled with PBS, and ultraviolet glue was applied to the coverslip edges.

*Preparation of C. elegans nematode samples*

Wild-type (N2) *C. elegans* nematodes were grown on nematode growth medium (NGM) plates seeded with Escherichia coli OP50-1 at 15 °C; 30–60 embryos, laid at 15 °C, were picked, transferred to new plates, and grown at 25 °C for the duration of the experiment. We determined the developmental stage of young adult nematodes using a light stereoscope. For BGM imaging, we first prepared two 0.25-mm-thick agar pads (5%) mixed with 10 mM sodium azide solution ($NaN_3$) to anesthetize the nematodes. Then, the agar pads were mounted on two 0.15-mm-thick round glass coverslips, 25 mm and 18 mm in diameter, and 10–15 young adult nematodes were sandwiched between the agar-padded coverslips. Ten microliters of an M9 contact buffer were added between the agar pads. To fix the entire sample and to avoid dehydration, ultraviolet glue was applied to the edge of the smaller coverslip, and a thin layer of Vaseline sealed the gap between the two coverslips.

*Preparation of NIH/3T3 cell samples*

NIH/3T3 cells (ATCC) were maintained using standard tissue culture procedures in a humidified incubator at 37°C with 5% $CO_2$ and atmospheric oxygen. They were grown in a

10 cm cell culture plate in 8 ml Dulbecco's Modified Eagle Medium (DMEM, Biological Industries) with 4.5 g/ml glucose. The DMEM was supplemented with 10% fetal bovine serum (FBS, Biological Industries) and 1% Antibiotic-Antimycotic (Anti-Anti). Cells were passaged by taking the media out and washing with 2 ml PBS (Biological Industries). Next, 1.5 ml of 0.25% trypsin and 0.002% EDTA solution (Biological Industries) were added to them. The cells were then incubated at 37°C with 5% $CO_2$ and atmospheric oxygen for 6 minutes until they were detached from the plate. 3.5 ml of DMEM was added to the detached cells and 0.5 ml of it was added into a new 10 cm plate with 8 ml DMEM. Cells were seeded at the amount of ~25,000–45,000 cells on a 25-mm diameter glass coverslip coated with Poly-D-Lysine (Merck) in a volume enclosed by adhesive spacers (Grace Bio-Labs SecureSeal). The cells were kept in the incubator for 24 hours in phenol red free DMEM medium (Biological Industries). For BGM imaging, cells were sandwiched between two glass coverslips (18 mm and 25 mm in diameter) with phenol-red-free DMEM medium. Vaseline ointment was used to ensure the complete sealing of the sample.

**Acknowledgments**

A.B. acknowledges the support of the Israel Science Foundation (grant no. 2576/21).

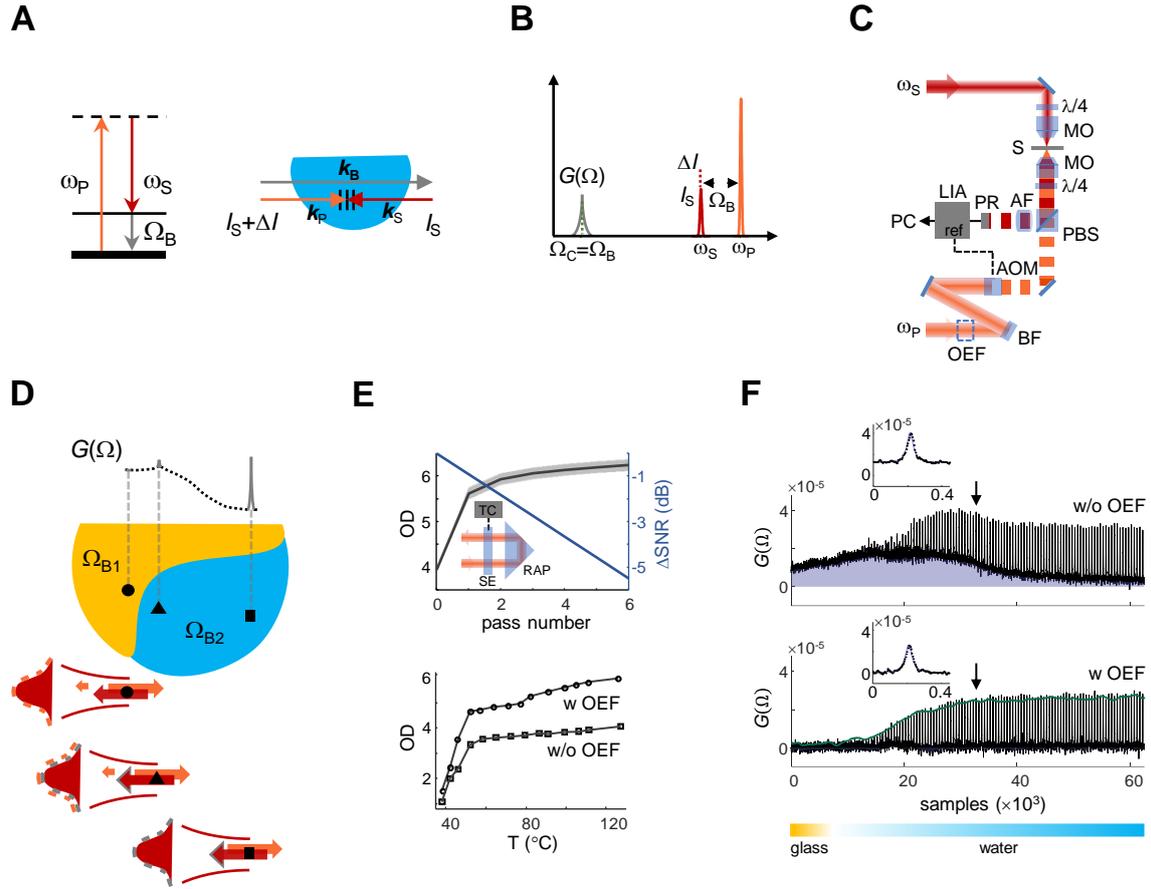

**Fig. 1.** SBS in optical mechanical microscopy. (**A**) Energy (left) and momentum (right) diagrams for stimulated Brillouin scattering (SBS; $\omega_P$ and $k_P$, pump frequency and wavevector; $\omega_S$ and $k_S$, Stokes frequency and wavevector; $\Omega_B$ and $k_B$, acoustic vibration frequency and wavevector in the sample). (**B**) Energy transfer between the pump and the Stokes beams in SBS. When the difference frequency between the excitation beams $\Omega = \omega_P - \omega_S$ is tuned over $\Omega_B$, the Stokes intensity $I_S$ increases by $\Delta I$ [indicated by the red dotted line and shown also in (**A**)]. $G(\Omega) = \Delta I/I_S$ is the stimulated Brillouin gain (SBG) at $\Omega$. In SBS microscopy, $G(\Omega)$ is obtained as a function of $\Omega$ by scanning $\omega_P - \omega_S(t)$ through $\Omega_B$ (gray line), whereas in BGM, $G(\Omega)$ is acquired at a single, mechanically contrasting frequency $\Omega_C$, e.g. $\Omega_B$ (green dotted line). (**C**) The counter-propagating Stokes and modulated pump beams are detuned by $\Omega$, circularly polarized by quarter waveplates ($\lambda/4$) and focused on a joint point in the sample (S) by the microscope objectives (MOs). $G(\Omega)$ is measured by detecting and demodulating the transmitted Stokes beam using a photoreceiver (PR) and a lock-in amplifier (LIA). Spectral data is collected and analyzed with a personal computer (PC). The difference frequency $\Omega$ is measured by an optical heterodyne detector and a frequency counter (not shown). The pump beam is locked to the D2 Rubidium-85 hyperfine resonance (not shown) and is modulated at 1.1 MHz with an acousto-optical modulator (AOM). The pump back-reflections are blocked by a Rubidium-85-based atomic filter (AF). The optical noise of the pump beam is filtered by a Bragg filter (BF). For BGM, an optimized etalon filter (OEF) is added to significantly suppress the pump back-reflection noise with minimum pump power loss. (**D**) A schematic displaying the baseline drift in $G(\Omega)$ (black dashed line) in a sample of two different mechanical constituents characterized by the acoustic vibration frequencies $\Omega_{B1}$ (orange) and $\Omega_{B2}$ (cyan). $G(\Omega)$ is measured over $\Omega_{B2}$ at three locations exhibiting a high baseline level due to significant pump back-reflection noise (light-red arrow and dashed line) from the interface (circle and triangle) and a low baseline level far from the interface (square). As a result, the SBG at $\Omega_C = \Omega_{B2}$ leads to erroneous identification of the different mechanical constituents in the sample. Stokes light and SBG light are shown in red and gray, respectively. (**E**) Optimized etalon filter (OEF). Simulations for optimizing the combination of etalon, Bragg, and atomic filters (top) showing the optical density (OD) and reduction of the signal-to-noise ratio (SNR) versus the number of passes through the etalon (solid black line, etalon reflectivity $R = 94\%$; gray shade, etalon reflectivity $R = 92–96\%$; inset; SE, solid etalon; RAP, right angle prism; TC, temperature controller). Measurements of the OD of the combination filter against temperature with and without the OEF (bottom; open circles, with OEF; open squares, without OEF). The OEF uses two passes through the etalon resulting in OD of ~6 and SNR reduction of ~1.3 dB at 110° C. (**F**) Measurement of $G(\Omega)$ versus samples along a glass−water interface without (top) and with (bottom) the OEF in the combination filter at 110° C. Individual SBG spectra (insets) and measurements of $G(\Omega_C)$ at $\Omega_C/2\pi = \Omega_B^{water}/2\pi = 5.03$ GHz (green line) are also shown. The pixel density in the $G(\Omega_C)$ measurements was twice than in the $G(\Omega)$ measurements.

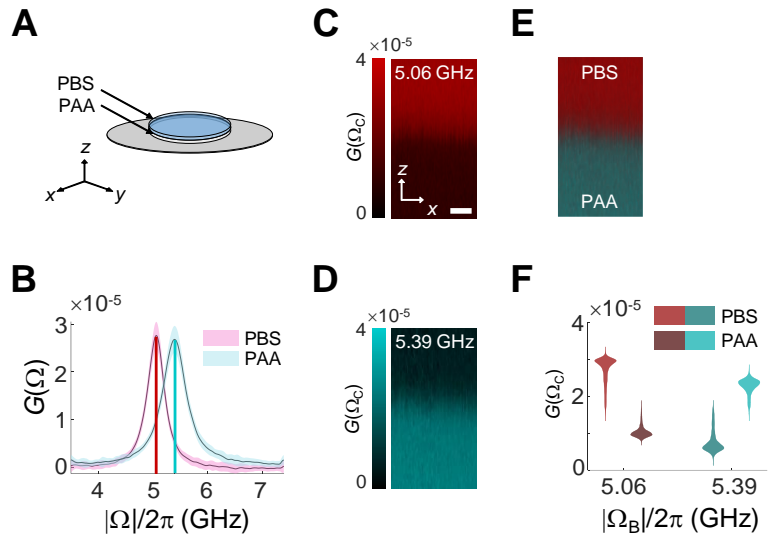

**Fig. 2.** BGM material imaging. (**A**) A double-layered material system consisting of a PBS layer on a PAA gel layer. (**B**) SBG spectra of PBS (magenta) and PAA (cyan), with peaks located at 5.06 GHz (red) and 5.39 GHz (cyan), respectively. (**C**) BGM PBS image in the material system shows PBS and PAA with positive and negative contrast, respectively. (**D**) BGM PAA image in the material system shows PAA and PBS with positive and negative contrast, respectively. (**E**) Dual-color overlay of the images (C) and (D). (**F**) Violin plots showing the extended distribution tails towards the PAA–PBS interface at 5.06 GHz and 5.39 GHz. The SBGs for the PBS and PAA layers at each acoustic vibration frequency are highly significantly different ($p$-value $< 2.23 \times 10^{-308}$) with visibility levels of 0.44 and 0.49, respectively. Visibility was computed as the ratio of difference-to-sum of the mean SBGs of the layers. Layers were segmented by the maximum between the SBGs at 5.06 GHz and 5.39 GHz in each pixel. Image size, ~80×210 pixels. Pixel size, 0.127×0.127 μm². Scale bar, 2.5 μm.

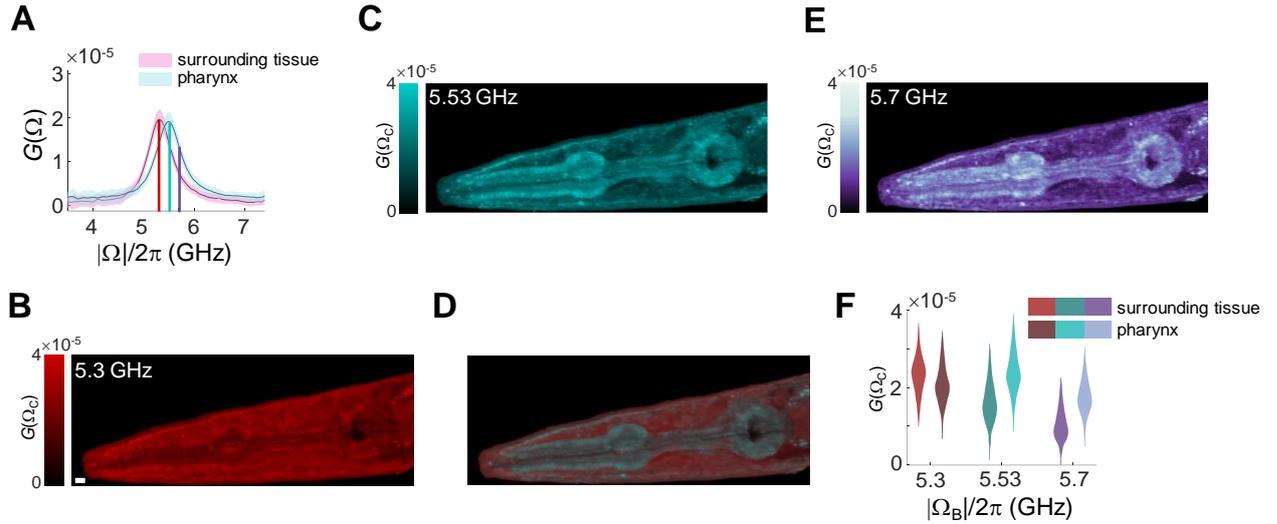

**Fig. 3.** BGM organism imaging in a living young adult *C. elegans* nematode. (**A**) SBG spectra of the surrounding pharyngeal tissue (magenta) and the pharynx (cyan), with peaks located at 5.3 GHz (red) and 5.53 GHz (cyan), respectively. (**B** and **C**) BGM images of the nematode's head at the acoustic vibration frequencies of the surrounding pharyngeal tissue (B) and the pharynx (C). The negative contrast in the pharynx channel (5.53 GHz) highlights open lumens in the nematode's head. (**D**) Dual-color overlay of the images (B) and (C). (**E**) BGM image of the nematode's head at the mechanically contrasting frequency of 5.7 GHz [indicated by a purple line in (A)] where mechanically induced changes in the Brillouin frequency shift are converted to a variation of the SBG. Spectral shift towards acoustic vibration frequencies below (above) 5.7 GHz reduces (increases) the SBG, resulting in darker (lighter) shades. (**F**) Violin plots of the surrounding pharyngeal tissue and the pharynx at 5.3 GHz, 5.53 GHz, and 5.7 GHz. Outliers were excluded for clarity. The SBGs for the surrounding pharyngeal tissue and the pharynx regions at each frequency channel are highly significantly different ($p$-value < $2.23 \times 10^{-308}$) with visibility levels of 0.06, 0.18, and 0.24, respectively. Visibility was computed as the ratio of difference-to-sum of the mean SBGs of the regions. Regions were segmented by the maximum between the SBGs at 5.3 GHz and 5.53 GHz in each pixel. Image size, ~1440×550 pixels. Pixel size, 0.127×0.127 μm$^2$. Scale bar, 5 μm.

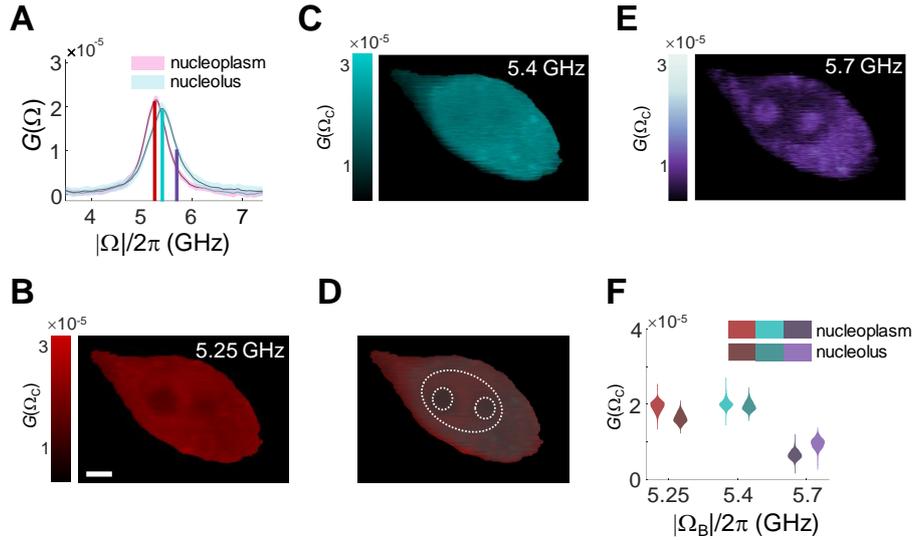

**Fig. 4.** BGM cell imaging in living NIH/3T3 fibroblasts. (**A**) SBG spectra of the nucleoplasm (magenta) and the nucleolus (cyan), with peaks positioned at 5.25 GHz (red) and 5.4 GHz (cyan), respectively. (**B** and **C**) BGM images of the cell at the acoustic vibration frequencies of the nucleoplasm (B) and the nucleolus (C). The negative contrast in the nucleoplasm channel (5.25 GHz) highlights the nucleoli. (**D**) Dual-color overlay of the images (B) and (C). Dashed lines outline the nucleoplasm and nucleoli. (**E**) BGM image of the cell at the mechanically contrasting frequency of 5.7 GHz [indicated by a purple line in (A)]. Spectral shift towards acoustic vibration frequencies below (above) 5.7 GHz reduces (increases) the SBG, resulting in darker (lighter) shades. (**F**) Violin plots of the nucleoplasm and the nucleoli at 5.25 GHz, 5.4 GHz, and 5.7 GHz. The SBGs for the nucleoplasm and the nucleoli regions at each frequency channel are highly significantly different ($p$-value $< 1.12 \times 10^{-46}$) with visibility levels of 0.09, 0.01, and 0.19, respectively. Visibility was computed as the ratio of difference-to-sum of the mean SBGs of the regions. Regions were segmented by the Otsu threshold applied to the histogram of the BGM image in (E). Image size, ~260×180 pixels. Pixel size, 0.127×0.127 μm$^2$. Scale bar, 5 μm.

Supplementary Material

# Brillouin Gain Microscopy

Roni Shaashoua,[1] Lir Kasuker,[2] Mor Kishner,[2] Tal Levy,[2] Barak Rotblat,[2,3] Anat Ben-Zvi[2,3] & Alberto Bilenca[1,4,*]

[1]Biomedical Engineering Department, Ben-Gurion University of the Negev, 1 Ben Gurion Blvd, Be'er-Sheva 8410501, Israel

[2]Department of Life Sciences, Ben-Gurion University of the Negev, 1 Ben Gurion Blvd, Be'er-Sheva 8410501, Israel

[3]The National Institute for Biotechnology in the Negev, Ben-Gurion University of the Negev, 1 Ben Gurion Blvd, Be'er-Sheva 8410501, Israel

[4]Ilse Katz Institute for Nanoscale Science and Technology, Ben-Gurion University of the Negev, 1 Ben Gurion Blvd, Be'er-Sheva 8410501, Israel

e-mail: *bilenca@bgu.ac.il

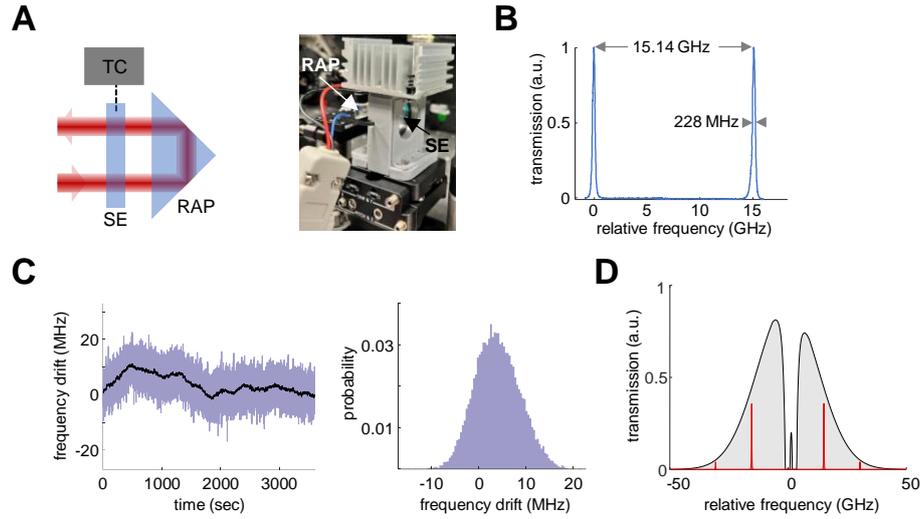

**Fig. S1.** Optimized etalon filter (OEF). (**A**) Schematic and photo of the OEF. (**B**) Transmission response of the OEF. The resultant two-pass finesse is ~66, which agrees well with the calculated value $F_1 / (2^{1/2} - 1)^{1/2} \cong 62$ where $F_1$ is the one-pass finesse measured to be ~40. This value of $F_1$ corresponds to a calculated etalon reflectivity of ~93%. (**C**) Frequency stability of the OEF. (**D**) Simulated transmission response of the Bragg−atomic filter (gray) and the optimized-etalon−Bragg−atomic filter (red) at 110° C with etalon reflectivity of 94%.

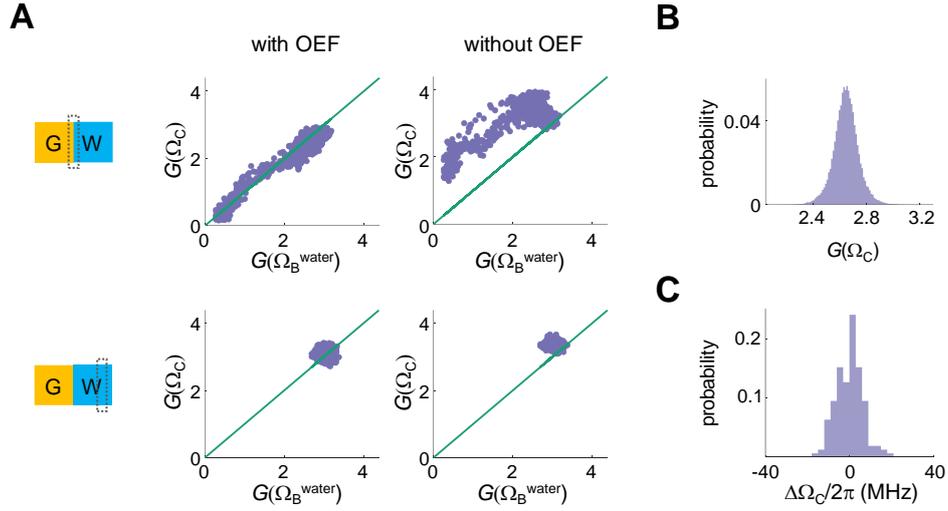

**Fig. S2.** Characterization of SBG measurements in water with and without the optimized etalon filter (OEF) at $\Omega_C/2\pi = \Omega_B^{water}/2\pi = 5.03$ GHz. (**A**) Correlation curves of $G(\Omega_C)$ versus $G(\Omega_B^{water})$ with (left column) and without (right column) the OEF near (top row) and far (bottom row) from the glass-water interface. $G(\Omega_C)$ was acquired by measuring the gain at $\Omega_C$. $G(\Omega_B^{water})$ was obtained by extracting the peak gain from the SBG spectrum measured with OEF following an additional algorithmic baseline drift correction. (**B**) Histogram of $G(\Omega_C)$. The mean and standard deviation are $2.64 \times 10^{-5}$ and $9.32 \times 10^{-7}$, respectively. $G(\Omega_C)$ was measured over ~6 sec with 100 μs temporal sampling. (**C**) Histogram of $\Delta\Omega_C/2\pi = (\Omega - \Omega_C)/2\pi$. The standard deviation is 6.2 MHz. $\Delta\Omega_C$ was measured over ~6 sec with a frequency counter. SBG values $G$ are $\times 10^{-5}$ in (A) and (B).

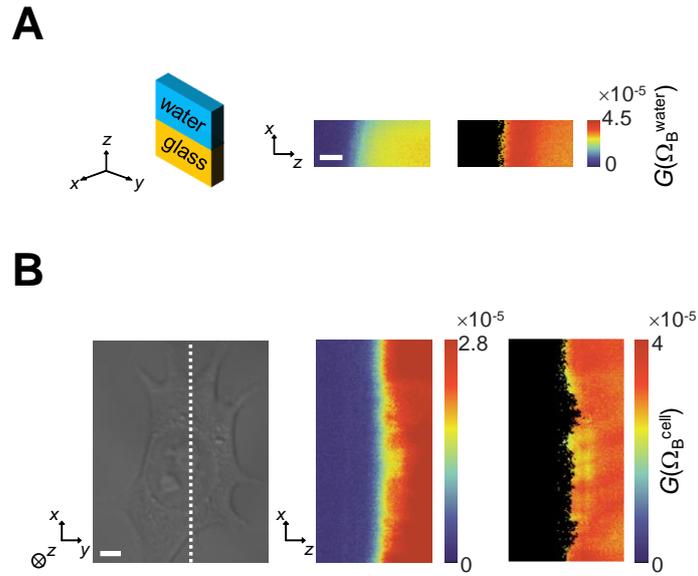

**Fig. S3.** Axial peak-SBG images across interfaces measured with and without the optimized etalon filter (OEF). (**A**) Glass-water interface with (left) and without (right) the OEF. The peak-SBG is significantly increased without OEF. (**B**) Glass-cell interface with (left) and without (right) the OEF. With the OEF, the peak SBG appears to be lower in the nucleolus than in the nucleoplasm since the nucleolus is stiffer than the nucleoplasm. Without the OEF, SBG data near the interface is not available. The peak-SBG was evaluated from the acquired SBG spectrum at each pixel in 20 ms with no algorithmic baseline drift correction. Scale, 5 μm.

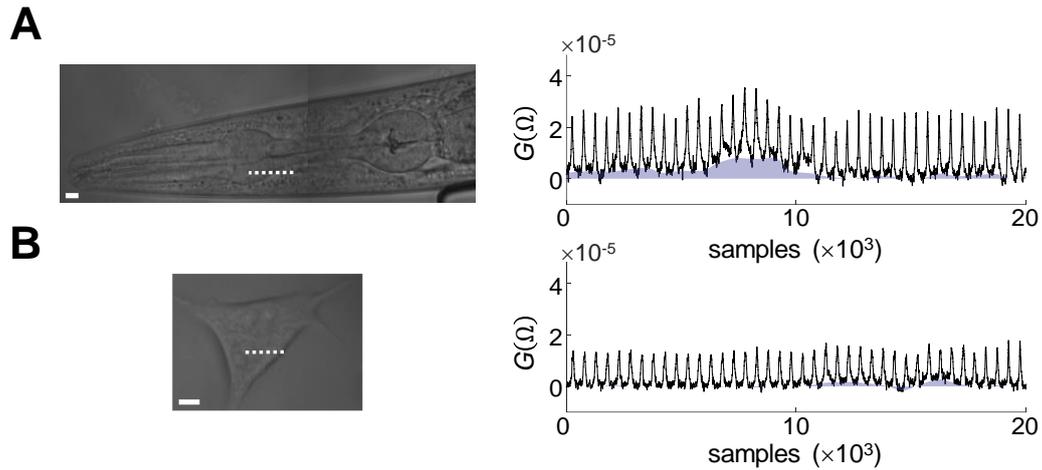

**Fig. S4.** Baseline drift correction in biological samples. (**A**) *C. elegans* nematode (left, brightfield image; right, SBG $G(\Omega)$ along the dashed white lines in the brightfield images). The deviation of the corrected baseline drift from the zero line is less than ~10% of the peak SBG outside the pharynx-surrounding tissue interface and is up to ~30% of the peak SBG across the interface. Etalons of different lengths in tandem could improve these deviations but at the expense of simplification. Scale, 5 μm. (**B**) NIH/3T3 cell (left, brightfield image; right, SBG $G(\Omega)$ along the dashed white lines in the brightfield images). The deviation of the corrected baseline drift from the zero line is up to ~20% of the peak SBG. Scale, 5 μm. Brightfield images were acquired concurrently with the SBG data.

**Note S1. Optimization of the etalon filter for baseline drift correction.**

To significantly improve the suppression of back-reflections of the pump beam into the path of the probe beam, the optical noise of the pump laser should be filtered in addition to the lasing mode. Tandem etalons or a multipass etalon can be used for this task, where the latter scheme greatly simplifies the filter implementation. While the objective of a multipass etalon is to significantly suppress the pump laser optical noise, the pump laser transmission should be kept high to guarantee an adequate signal-to-noise ratio. For the given etalon reflectivity ($R = 94\pm2\%$), we optimized using MATLAB the number of passes to achieve optical density and transmission of the multipass-etalon–Bragg–atomic filter of ~$10^{-6}$ and >80%, respectively, yielding a back-reflected pump-to-Brillouin signal ratio <0.1 and a signal-to-noise ratio degradation <2 dB assuming SBG of $2\times10^{-5}$, detected probe and back-reflected pump powers of 5.5 mW and 4% × 222 $\cong$ 8.9 mW, respectively, and pump laser optical noise suppression of 39.3 dB (Toptica) [top panel of Fig. 1E and Fig. S1D]. The multipass etalon was modeled as $T_{max}^m / \left[1 + F\sin^2(\delta/2)\right]^m$ with measured one-pass maximum transmission of $T_{max} = 90\%$. Also, $m$ represents the pass number, $F = 4 \times R / (1 - R)^2$, and $\delta = 4\times\pi / \lambda \times n \times l$ where $\lambda \cong 780.24$ nm, $n = 1.453661$, $l = 6.7395$ mm (LightMachinery). The Bragg filter was approximated by a Gaussian function with 30 GHz full-width at half-maximum and 96% diffraction efficiency (OptiGrate), and the atomic filter was modeled at a temperature of 110° C[26].